\documentclass[sigconf]{acmart}
\AtBeginDocument{%
  \providecommand\BibTeX{{%
    \normalfont B\kern-0.5em{\scshape i\kern-0.25em b}\kern-0.8em\TeX}}}

\copyrightyear{2024} 
\acmYear{2024} 
\setcopyright{rightsretained} 
\acmConference[CHI EA '24]{Extended Abstracts of the CHI Conference on Human Factors in Computing Systems}{May 11--16, 2024}{Honolulu, HI, USA}
\acmBooktitle{Extended Abstracts of the CHI Conference on Human Factors in Computing Systems (CHI EA '24), May 11--16, 2024, Honolulu, HI, USA}
\acmDOI{10.1145/3613905.3644063}
\acmISBN{979-8-4007-0331-7/24/05}

\usepackage{hyperref}
\usepackage[utf8]{inputenc}
\begin{document}

\title[``That's Not Good Science!'']{``That's Not Good Science!'': An Argument for the Thoughtful Use of Formative Situations in Research through Design}

\author{Raquel B Robinson}
\email{raqr@itu.dk}
\orcid{0000-0003-4933-7168}
\affiliation{%
  \institution{IT University of Copenhagen}
  \city{Copenhagen}
  \country{Denmark}
}

\author{Anya Osborne}
\email{akolesni@ucsc.edu}
\orcid{0000-0002-5506-623X}
\affiliation{%
  \institution{University of California, Santa Cruz}
  \city{Santa Cruz}
  \state{California}
  \country{USA}
}

\author{Chen Ji}
\email{cji40@ucsc.edu}
\orcid{0000-0001-6554-0217}
\affiliation{%
  \institution{University of California, Santa Cruz}
  \city{Santa Cruz}
  \state{California}
  \country{USA}
}

\author{James Collin Fey}
\email{jfey@ucsc.edu}
\orcid{0000-0002-5033-0134}
\affiliation{%
  \institution{University of California, Santa Cruz}
  \city{Santa Cruz}
  \state{California}
  \country{USA}
}

\author{Ella Dagan}
\email{ella@ucsc.edu}
\orcid{0000-0003-1032-7018}
\affiliation{%
  \institution{University of California, Santa Cruz}
  \city{Santa Cruz}
  \state{California}
  \country{USA}
}

\author{Katherine Isbister}
\email{kisbiste@ucsc.edu}
\orcid{0000-0003-2459-4045}
\affiliation{%
  \institution{University of California, Santa Cruz}
  \city{Santa Cruz}
  \state{California}
  \country{USA}
}

\renewcommand{\shortauthors}{Robinson et al.}

\begin{abstract}
Most currently accepted approaches to evaluating Research through Design (RtD) presume that design prototypes are finalized and ready for robust testing in laboratory or in-the-wild settings. However, it is also valuable to assess designs at intermediate phases with mid-fidelity prototypes, not just to inform an ongoing design process, but also to glean knowledge of broader use to the research community. We propose 'formative situations' as a frame for examining mid-fidelity prototypes-in-process in this way. We articulate a set of criteria to help the community better assess the rigor of formative situations, in the service of opening conversation about establishing formative situations as a valuable contribution type within the RtD community.
\end{abstract}





\maketitle

\section{Introduction}
\begin{figure*}
  \includegraphics[width=\textwidth]{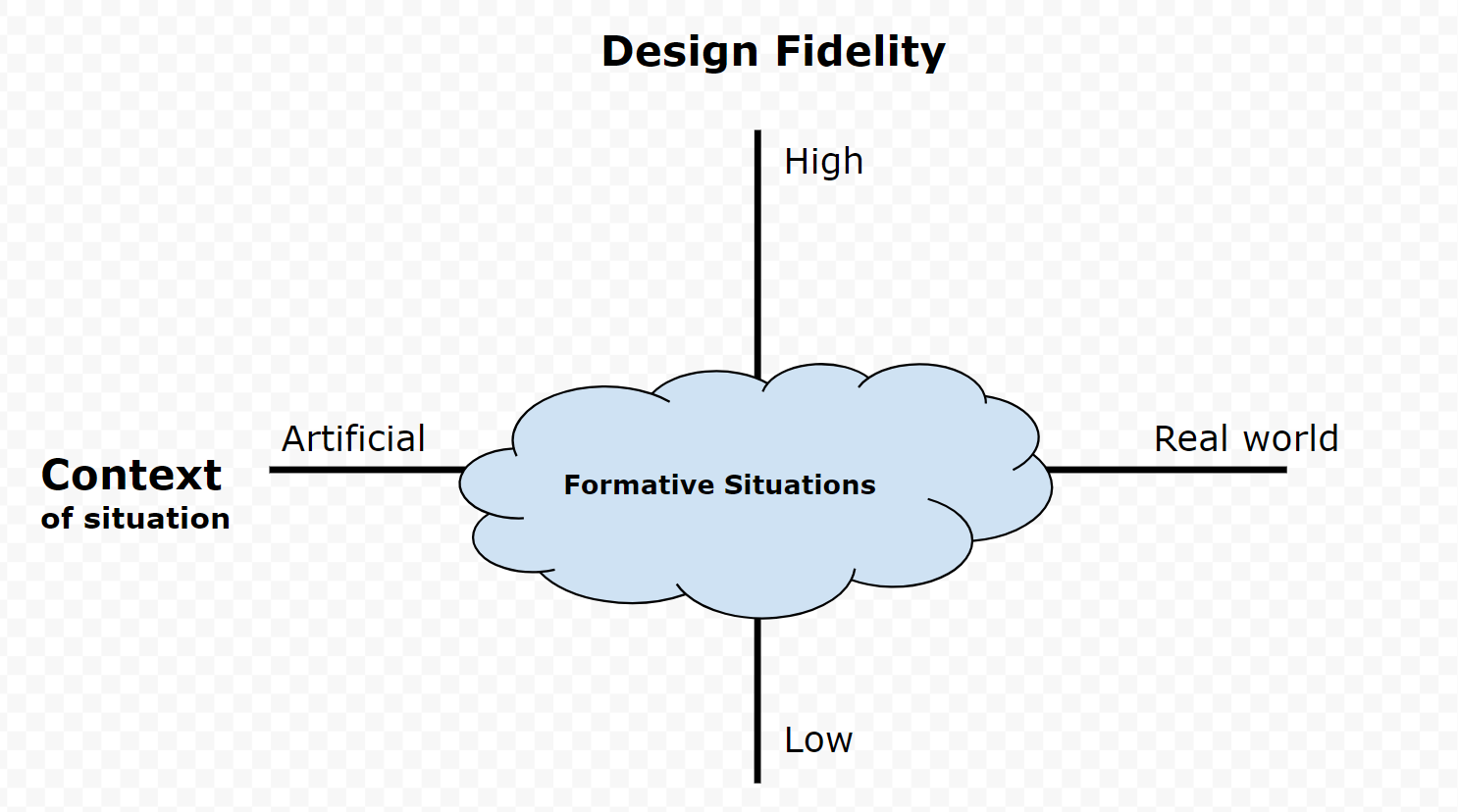}
  \caption{Formative Situations in RtD}
  \Description{Graph of two axes: y-axis is design fidelity from low to high. X-axis is context of situation from artificial to real world.}
  \label{fig:teaser}
\end{figure*}
We begin by posing a question to the HCI community for consideration: why do we as a research community seem to value summative evaluation over formative evaluation? We posit two primary reasons: i) researchers don’t yet understand the extensible and useful knowledge gained from formative evaluations, and thus seek to make a summative empirical contribution when conducting evaluations \cite{MacKenzie2013Human-ComputerPerspective, Wobbrock2016ResearchInteraction} rather than formulating the intermediate level knowledge that formative evaluations provide and ii) reviewers conflate formative and summative evaluations erroneously, holding formative evaluations to inappropriate standards. This conflation may be partly because there is not yet an established approach for evaluating artifacts at a more formative phase of the continuum of design.  

To address this gap, we present `\textit{formative situations}'. What we mean by a formative situation is not a preliminary evaluation of a design concept for a work in progress piece, but rather, a deep examination of a mid-fidelity prototype that results in valuable knowledge for the community as such.


Formative evaluations - or evaluations that are used to provide insights which inform and guide the design process - are a foundational concept within the field of HCI. Textbooks often used to teach HCI courses such as Interaction Design: Beyond Human-Computer Interaction \cite{Rogers2002InteractionInteraction} and Human-Computer Interaction \cite{Dix2003Human-ComputerEdition} explain both \textit{formative} and \textit{summative} evaluations as fundamental contributions of knowledge to HCI. However, while the concept of formative evaluations may be touched upon, HCI course curriculums are heavily centered around trying out primarily summative forms of evaluations, for example field deployments where the prototype is tested with the target population \cite{Siek2014FieldContext} and lab studies where one can control many aspects of the situation \cite{MacKenzie2013Human-ComputerPerspective}. This problem extends outside of the classroom as well: HCI publication venues heavily emphasize and value summative forms of evaluation that aim to understand how well a design performs, usually leveraging controlled experiments and field studies, and sidelining formative evaluations entirely. In 2016, ``empirical" \footnote{see next section for our account of the word empirical} research made up the largest portion of contribution types at the ACM Conference on Human Factors in Computing Systems (CHI) \cite{Wobbrock2016ResearchInteraction}---either in-lab or from the field, which are evaluated on the ``importance of their findings and on the soundness of their methods" \cite{Wobbrock2016ResearchInteraction}. 

But is this appropriate? In fact, we argue that much of the design work done in the field of HCI seems to fall more into the category of mid-fidelity prototype, and thus might more appropriately be formatively assessed. Fällman notes that design researchers in the field of HCI usually only go as far as to develop functioning prototypes \cite{Fallman2007WhyDiscipline}. He states that, "Many of the prototypes that researchers develop are, too, anything but convincing products. They may be wholly or partly faked; if implemented, they may be unstable and lack some expected functionality" \cite{Fallman2007WhyDiscipline}. Despite this, much of our design work is still evaluated in a summative situation at one of the two extremes (i.e., in a lab setting or in the wild with real users) rather than in a more formative manner---see on the x-axis of Figure \ref{fig:teaser}.

What if instead, we evaluated our mid-fidelity prototypes within the context of a situation of equal proportionality to the prototype? Figure \ref{fig:teaser} shows the design fidelity of the artifact on the y-axis, and the context of the situation on the x-axis (from artificial to real world). We see formative situations as existing at the middle of this continuum - part artificial so the designer is still able to address any functionality issues that may arise with the prototype \cite{Siek2014FieldContext} and part situated in a real world context. 

Formative situations as we envision them could employ some previously validated measures and methods that we expect from summative evaluations, but a formative situation could also include provisional measures as suited to the situation, and generally take a deeply adaptable and flexible approach, based on the fidelity of the artifact and research goals (shown in Table \ref{tab:methods}). 

We would argue that many HCI projects at CHI in recent years might have benefited from undergoing a longer design cycle that included one or more formative evaluation stages, each of which could result in valuable, extensible knowledge for the larger community. Privileging and requiring summative evaluation for publication puts immense pressure on researchers to put their prototypes through summative evaluation as quickly as possible - often before the design is ready to be evaluated in such a way. We propose a paradigm shift in the community of Research through Design: instead of \textit{prematurely summatively evaluating} design prototypes, consider formative evaluations as the first publishable stage of the research knowledge generation process. This shift could help take the pressure off the community to conduct extensive, large scale evaluations of unfinished, mid-range prototypes of systems, freeing up time to focus future work on design refinements that lead to summative testing where warranted, or on sharing formative design learnings with those who could further develop them outside academia into products for commercial or non-profit use.

Allowing for shareable research contributions from formative evaluations along the way to final designs worthy of summative evaluation could free our community to build more flexibly and extensively upon one another's work, interweaving the expertise of research labs that excel in different stages of the design process. One lab might generate many design insights from formative work, another might bring some of these ideas to rich fruition in high fidelity prototyped designs rigourously tested, perhaps yet another group finishes the cycle by producing what Odom et al. characterize as a ``research product'' that is robust and widely distributed \cite{Odom2016FromProduct}. At any point, the actionable design insights could be passed over to design practitioners \cite{VanBerkel2023ImplicationsResearch} to make use of in developing commercial products. We believe it is possible that an excess of rough and uncredible `finished' prototypes in our research literature could actually be preventing us from passing along workable design knowledge that experts in commercial development and practice can take much further in a better way.

\begin{table*}[h]
\caption{Comparison of summative and formative RtD situations for evaluation}
\label{tab:methods}
\small
\begin{tabular}{|p{1.3cm}|p{5cm}|p{5cm}|} 
\cline{1-3}
         & \textbf{Formative}  & \textbf{Summative}                                              \\ \cline{1-3}
         
\textbf{Research Goals} & Inform design process and glean shareable design knowledge & Assess overall experience of using system  \\ \cline{1-3}
\textbf{Measures} & Use of provisional measures suited to the situation AND previously validated measures & Use of previously validated measures                                          \\ \cline{1-3}
\textbf{Researcher} & Integrated into the situation &  Limited involvement to minimize biases  \\ \cline{1-3}
\textbf{Setting} & Settings that are semi-natural/semi-wild are accepted and cultivated & Employs experimental design strategies or related best practices (lab study, field deployment)  \\ \cline{1-3}
\textbf{Methods} & Provisional and adapted methods, modified to meet the RtD objectives &  Employs methods that are well-established  \\ \cline{1-3}
\textbf{Benefits} & Transparency of research process; greater potential for translation to industry contexts & Generalizations or comprehensive look into design space or artifact  \\ \cline{1-3}

\end{tabular}
\end{table*} 

\section{Background}
What is ``rigor'' in scientific work? This question has been a widely debated topic in the field of HCI for decades \cite{Stolterman2008TheResearch, Olson2014WaysHCI}. HCI is a multi- and interdisciplinary field, consisting of practices and research methods applied from domains including computer science, psychology, design, and sociology among others \cite{LazarFengHochheiser17}. Thus, there exists a wide variety of approaches to conducting research within the field. Each particular sub-discipline (e.g., Visualization, Health, Games and Play, Design) has their own practices and established methods, and therefore discussions of what constitutes rigorous and valid work \cite{Gaver2012WhatDesign}. Similarly, the kind of knowledge \cite{Wobbrock2016ResearchInteraction} and ways of knowing \cite{Olson2014WaysHCI, Laurel2003DesignPerspectives} that are generated from each is widely varied. Contributions range from higher level discussions of theory and meta-reflections (e.g., how ethnography research should be evaluated in HCI \cite{Dourish2006ImplicationsDesign}) to very specific use cases (e.g., a physiologically-adaptive video game to support individuals with emotion regulation \cite{Lobel2016Nevermind:Game}). This variance in research approach is similarly paralleled by a wide range of accepted methods for evaluating research artifacts. Some fields within HCI use experimental psychology techniques and the scientific method to test hypotheses and address larger research questions \cite{Schon1987EducatingProfessions, Moher1996AssessingDirections/i, Campbell2018b}. Researchers using these methods often evaluate their artifacts in a highly controlled situation within a laboratory environment \cite{MacKenzie2013Human-ComputerPerspective}. On the other hand, design-focused research approaches within HCI can be characterized in many different ways - while the former approach is research-question driven with the artifact existing as supplement, design researchers are often said to be driven by the designed artifact itself. Wobbrock et al. state that “Whereas empirical contributions arise from descriptive discovery driven activities (science), artifact contributions arise from generative design-driven activities (invention) \cite{Wobbrock2016ResearchInteraction}. Design is a highly situated and contextual practice, in which the "success" of the design is dependent on many factors including the individual or social context with which its deployed. Design research questions are quite complex - addressing problems that are said to be unsolvable, or "wicked" \cite{Rittel1973DilemmasPlanning}. Design is often also considered an individual and intuitive practice - where the designer themself is also an important part of determining success of the designed artifact \cite{Schon2017ThePractitioner, Prochner2022QualityEnhancement}. Debate has risen over the years as to whether or not design is a science, an art, or somewhere in the middle \cite{Cross2001DesignerlyScience, Gaver2012WhatDesign}. In 2003 Fallman published a paper on design-oriented human-computer interaction, discussing three competing accounts of conceptualizing design research in HCI. The three competing perspectives are the \textit{romantic account} which focuses on the value and taste of the designer, the \textit{conservative account} which follows a more structured approach most akin to science, and the \textit{pragmatic account} which focuses on the engagement with the particular design situation. This, in turn has caused much debate about how to evaluate the work that comes out of design-oriented research within HCI.

There is a rich history of many kinds of design work at CHI (e.g., critical design, design fiction, and speculative design \cite{Dunne2006HertzianDesign, Dunne2013SpeculativeDreaming, Baumer2020EvaluatingJob} to name a few), however our perspective in this paper is aimed toward the area of \textit{Research through Design}. Research through Design is a phrase coined by Frayling et al. in 1993 \cite{Frayling1993ResearchDesign}, which over the years has solidified as an established approach to research within the field \cite{Zimmerman2007ResearchHCI}. RtD is an approach to conducting scholarly research ``that employs the methods, practices, and processes of design practice with the intention of generating new knowledge''. \cite{Zimmerman2014ResearchHCI} Although RtD has become a widespread approach, considerable confusion still exists concerning the criteria by which designed artifacts and the knowledge generated from their creation should be evaluated. Some argue that the field of Research through Design needs to have an agreed upon set of standards for practice, evaluation, and outcome \cite{Zimmerman2007ResearchHCI} to show progress and results, while others argue that having such standards might hinder progress and the field may "develop not only through increasing agreement, but also through discursiveness and elaboration" \cite{Gaver2012WhatDesign}. Gaver and Bowers note that "Methodological frameworks promise rigor but jeopardize the possibility for designers to invent ad hoc approaches, or draw inspiration from unorthodox sources, or take inexplicable imaginative leaps---all forms of a productive indiscipline that we see as integral to design practice." \cite{Gaver2012AnnotatedPortfolios} We agree with this point that one of the strengths of design research and practice is in the emergent and novel output \cite{Gaver2022EmergenceResearch} it generates, however we also take the stance that there is a heavy reliance on certain regularized and accepted methods of evaluation that are entirely summative. In the current HCI landscape, the most widely accepted methods of evaluating an artifact include situations that are either empirically derived \cite{Schon1987EducatingProfessions}) or a situation that is fully integrated into a real world context within the ecosystem of the target community \cite{Rogers2011InteractionTheory, Chamberlain2012ResearchDevelopment}. However, the design process is highly iterative, and consists of many different phases. It stands to reason that there is valuable, extensible knowledge somewhere between contributions that are \textit{unfinished, works in progress} and \textit{summative evaluations}.

\subsection{Defining Contributions in HCI}
More broadly, HCI as a discipline is interested in the generation of new knowledge; the purpose and outcomes of that knowledge, however, vary according to the particular sub-discipline and project goals. In a 2016 report, Wobbrock et al. designate particular kinds of contributions within the field of HCI, and how they differ in the knowledge produced and the ways of evaluation \cite{Wobbrock2016ResearchInteraction}. Two of the top contribution types noted were \textit{Empirical} and \textit{Artifact} contributions. While empirical contributions in HCI are said also involve designing an artifact or system, the distinct contribution of this kind of work lies in the empirical study itself which highlights how people interact with the system. On the other hand, Research through Design is often responsible for many of the \textit{artifact contributions} in HCI. This type of contribution usually includes a designed artifact which is generated by a particular design process, as well as some kind of evaluation to say whether or not the artifact has achieved the intended goal set by the designer and target user group. The context in which the evaluation takes place can range on a spectrum from fully artificially constructed (e.g., lab study \cite{Schon1987EducatingProfessions}) to fully socially viable and integrated into the target community with real world use cases (e.g., in-the-wild \cite{Rogers2017ResearchWild, Chamberlain2012ResearchDevelopment}). While the former usually leverages studies that are empirically motivated and offers high internal validity (i.e., cause-and-effect relationships between elements of the research \cite{Prochner2022QualityEnhancement}), the latter helps understand how the artifact will fit into real social ecosystems (ecological validity \cite{Lewkowicz2001TheInvalid}) and generalizability outside the particular context (external validity) \cite{GubaYvonn1994CompetingResearch, Campbell1966ExperimentalResearch}. 


Oulasvirta \& Hornbaek (building off a philosophy proposed by \cite{Laudan1977ProgressGrowth} in 1977) - state that there are three types of HCI contributions: \textit{empirical, conceptual, and constructive}. \textit{Empirical} research is "aimed at creating or elaborating descriptions of real-world phenomena related to human use of computing", \textit{conceptual} research is "aimed at explaining previously unconnected phenomena occurring in interaction", and \textit{constructive} research is "aimed at producing understanding about the construction of an interactive artefact for some purpose in human use of computing \cite{Oulasvirta2016HCIProblem-solving}. However, both Wobbrock et al. and Oulasvirta et al. note that empirical contributions made up the largest portion of contributions to the ACM CHI Conference on Human Factors in Computing (the leading conference in Human-Computer Interaction) in 2015 \cite{Oulasvirta2016HCIProblem-solving} and over half the contribution types in 2016 \cite{Wobbrock2016ResearchInteraction}. Conference venues such as CHI follow double-blind, peer-reviewed processes \cite{CHI2024}, meaning the community itself is responsible for deciding whether or not the rigor of the work is acceptable. The abundance of empirical work may perpetuate a reinforcing feedback loop; as more of this kind of work is output and early researchers enter the field and see an abundance of this same kind of work, they assume this is the gold standard.

Evaluations in the context of HCI take inspiration from varied fields. `Empirical' studies are inspired in a broad sense by the scientific method and more narrowly by a paradigm common in experimental psychology \cite{MacKenzie2013Human-ComputerPerspective}. In experimental psychology, causality is a vital consideration: does x intervention cause y behavior or emotion. Laboratory experiments aim to control for all other variables than the one of interest, and practitioners aim to use previously validated methods and measures, as shown in table \ref{tab:methods}. In medical fields, researchers run carefully designed clinical trials for testing new drugs and treatment protocols, that also aim to isolate causality by controlling variables other than the intervention itself \cite{Moher1996AssessingDirections/i}. Though HCI rarely tests artifacts with as large a number of participants as a drug trial, nor uses techniques like blinding which condition a subject is in from the experimenter, empirical HCI practitioners in some sense aim to adhere to the standards set by these fields \cite{MacKenzie2013Human-ComputerPerspective}. 

Another key tradition in summative evaluation is field studies. There are many approaches to field studies used in HCI \cite{Siek2014FieldContext}, with the general purpose being to try the artifact out within the environmental ecosystem of the target user group \cite{Chamberlain2012ResearchDevelopment, Rogers2017ResearchWild}. Field studies can achieve high external validity because they are conducted within end user contexts. However, many factors can’t be controlled in a field study, so it is impossible to establish clear causality of system effects. An interesting sub-genre of field studies is technology probes. Technology probes get closer to what we mean by formative situations, in that the goal is not to summatively test a specific prototype, but rather, ``understanding the needs and desires of users in a real-world setting, the engineering goal of field-testing the technology, and the design goal of inspiring users and researchers to think about new technologies" \cite{HutchinsonTechnologyFamilies}.

While empirical work makes up the largest number of contributions within HCI, \textit{artifact} focused work also accounts for much of the work generated by the field \cite{Fallman2003Design-orientedInteraction}. In 2003, Fallman discusses HCI as a design-focused discipline - and that there needs to be a distinction in design-oriented research (the generation of knowledge) and research-oriented design (the generation of artifacts), and the criteria on which we judge them. However, much confusion still exists around how design-focused HCI work should be judged as "rigorous" and assessed within the community, as it does not and sometimes should not \cite{Gaver2012WhatDesign} conform to empirical standards. On one hand, some researchers feel that there should be some kind of detailed description of the design process and method of evaluation (i.e., transparency) \cite{Zimmerman2007ResearchHCI}, and defensible,  or grounded empirically, analytically, and theoretically \cite{Hook2012StrongResearch}. Prochner et al. simiarly highlight numerous difficulties in evaluating RtD work and propose a possible set of research quality indicators motivated by empirical categories (e.g., Traceability, Impartiality), and discussing how these indicators may be used by designers to more thoughtfully describe and position their work \cite{Prochner2022QualityEnhancement}. Gaver et al. highlight this community standard as well: "clearly articulated intentions at the outset of a project are a major resource used in making judgements about the ‘rigour’ of the research project and the outcomes it produces." \cite{Gaver2022EmergenceResearch}. However, Gaver argues that one of the strengths of design is in its emergence (or, “something arising out of ongoing activity, enacted rather than predetermined”), and thus proposes a set of criteria for how to both discuss and assess the contribution. In terms of how to assess the contribution, he proposes three criteria: 1. \textit{Research objectives are provisional and not contractual} - or ``does the description of the artifact offer a launching point for exploration"? 2. \textit{Assess outputs on their own term}s - or ``look to the output as grounds for success" 3. \textit{Value agility and responsiveness} - or ``are new ideas allowing a promising route forward"? Building off these proposed criteria as a basis, we propose our own set of criteria in this paper particularly for \textit{formative} research through design situations. 

We see the kind of formative contribution we propose as empirical as well - work that should be evaluated on its own merits. The word `empirical' is sometimes used loosely and interchangeably within HCI work to mean use of the scientific method. Empirical work is defined by Mackenzie et al. as ``experimentation to discover and interpret facts, revise theories and laws'' and ``is capable of being verified or disproved by observation or experiment'' \cite{MacKenzie2013Human-ComputerPerspective}. However, empirical is defined in the Cambridge dictionary as ``based on what is experienced or seen rather than on theory'' \footnote{\url{https://dictionary.cambridge.org/dictionary/english/empirical}}. We would like to establish \textit{our} understanding and use of this word for the purposes of this paper. We see empiricism as making observations in the world and adjusting to those observations, or in line with Oulasvirta et al.'s definition: ``aimed at creating or elaborating descriptions of real-world phenomena related to human use of computing'' \cite{Oulasvirta2016HCIProblem-solving}. Thus, we argue for the perspective that design research \textit{is} in fact empirical research, yet not making use of the scientific method. We see this kind of formative research as both rigorous and empirical - the ultimate goal to evaluate the artifact within a situation that elicits rich, nuanced feedback from participants. Our goal in the following sections is to educate reviewers on these criteria for evaluation, to allow these formative situations to be more accepted and established within the HCI community in future.

\section{Proposed criteria for the assessment of formative situations}
Here we articulate a set of criteria for designing a formative situation for evaluating a given design artifact, that can aid reviewers in assessing the study's contribution.  

These criteria arose from extended discussion within our lab group that works at the intersection of HCI and games--we authors collectively have been both practicing designers and researchers within HCI for many years. Some time ago, our lab realized we have actually been crafting formative situations for years. This work often goes through multiple review cycles, because on first review it is received by the community as not rigorous enough, or not conforming to pre-existing standards, before eventually finding publication homes (e.g. \cite{Dagan2020FlippoCompanion}, \cite{Dagan2021SynergisticInteraction}, \cite{Isbister2022DesignRegulation}, and \cite{Ji2022ARFidgeting}). In discussion with other researchers in the field who had similar challenges, we realized there seemed to be a collective \textit{methodological gap} here that needed to be articulated. We propose the following criteria as a launching point for discussion and further refinement. 

The primary goal in designing this kind of formative situation is to elicit participants to give rich, subjective feedback about the design of the prototype, not just for the purpose of informing future design iterations, but also for gleaning extensible and shareable design knowledge. We present this set of criteria to open up the dialogue at CHI about the acceptance of more formative work.

\begin{table}[htbp!]
\begin{tabular}{|p{1.2cm}|p{7cm}|}
\cline{1-2}
\textbf{Criterion Label} & \textbf{Description of Criterion}                             \\ \cline{1-2}
C1                       & Well fitted to the RtD questions                                                           \\ \cline{1-2}
C2                       & Draws upon related best practices                                 \\ \cline{1-2}
C3                       & Of appropriate weight compared to the provisionality of prototype                                                          \\ \cline{1-2}
C4                       & Allows designer to give integrative feedback                                               \\ \cline{1-2}
C5                       & Allows participants an active role in the process                                            \\ \cline{1-2}
\end{tabular}
\label{tab:cri}
\caption{Criterion for the Assessment of Formative Situations}
\end{table}

\subsection{Criteria}
\hspace{\parindent} \textit{(C1) Well fitted to the RtD questions}: Criteria for creation of the situation should be detailed and clear, and guided by the design research questions. The formative situation should be clearly described and all those who engage with it experience something similar (to allow the researcher(s) to contextualize feedback received). 

\textit{(C2) Draws upon related best practices}: Clearly outline the design process and self-articulate the particular design phase of the prototype for the purposes of transparency. Detail how the chosen formative situation is of appropriate weight in comparison. Describe which method you adapted the situation from (if there are any). As a reviewer, do not assume each contribution is at the same stage of the design process. 

\textit{(C3) Of appropriate weight in proportion to the provisionality of the prototype}: As outlined in Figure \ref{fig:teaser}, the situation for study should match the provisionality/design resolution of the artifact (i.e., if a prototype needs extensive support it may not yet be viable to bring into an 'the wild' situation)). 

\textit{(C4) Allows designer an appropriate position within the process}: Formative situations can more fluidly involve the designer-researcher, in contrast to the experimental paradigm where it is important to remove potential bias from the experimenter. In a formative situation, the researcher might adapt the situation in real-time to adjust for emergent responses and reactions from the participants, clarify where needed, and/or babysit sometimes fragile prototypes. 

\textit{(C5) Allows participants an active role in the process}: Formative situations can offer opportunity for resistance, and discussion from prospective users may be necessary as to understand the contours and rough edges of a design. 

Evaluation situations consist of inherent power dynamics; an unspoken social contract between the designer-researcher and the participants. In experimental psychology, participants are often called ``human subjects'', as a linguistic nod to the fact that the participants are \textit{subjects} of the research itself. Often, in experimental studies, the experimenter knows more than the participants (i.e., regarding specific research goals, how the system is expected to work), and should not notify the participants of these things in order not to bias their opinions of the tool. We propose that formative situations take a different approach - one that is reflexive (the process of a continual internal dialogue and critical self-evaluation of researcher’s positionality as well as active acknowledgement and explicit recognition that this position may affect the research process and outcome \cite{Pillow2003ConfessionResearch, Boehner:2007:EMM:1225302.1225522, Berger2013NowResearch}), egalitarian, and self critical - involving no deceit, helping to minimize the power dynamics inherent to the situation. In these situations, participants are not \textit{subjects} to be studied, rather they are treated as equals. The goal of the designer-researcher is to help participants understand the technology on a deep enough level to be able to give feedback. This often involves directly telling participants the goals of the study, the research questions, and the expected output of the system (which are things often frowned upon in experimental research studies). 

\subsection{What if these criteria are not met?}
While we establish a broad range of what can be considered `formative' research, we also need to understand what constitutes a formative situation that does \textit{not} satisfy the proposed criteria. Research which meets the proposed criteria and demonstrates validity through the points described below provides a significant knowledge contribution to the community. Below we describe how to assess the situation in terms of its ``\textit{rigor}''.

Empirical studies usually judge validity on the basis of \textit{replicability}, \textit{internal and external validity}, \textit{objectivity}, and \textit{reliability} \cite{GubaYvonn1994CompetingResearch}. We draw upon and adapt the principles behind these criteria, toward providing rigor without overly limiting the possibilities for emergent design outcomes. The goal with the below points aid in the \textit{reviewability} of formative situations, to help reviewers understand if the criteria have been met, and what we see as an ``insufficient'' description of the formative situation.
\begin{itemize}
    \item \textit{No Transparency}: Enough detail should be provided about the research and formative situation to enhance the transparency of research  and overall design process. If the particular stage of the design process or fidelity of the prototype is not outlined, then the formative situation can't be adequately assessed in relation to the provisionality of the artifact. 
    \item \textit{Not well aligned with real world situations and settings}: In this case, the formative situation described does not draw well from existing everyday experience, whether that is a particular context, or a particular social or personal situation that can be somehow evoked in a lab setting. Such real world alignment could be achieved by engaging with participants that might use such a system were it developed, crafting social situations that act as a microcosm for situations the technology might afford in everyday life, or having the engagement with the prototype take place in a more naturalistic environment such as a home or office setting. The aim here is to err toward \textit{external} and \textit{ecological} validity. 
    \item \textit{Too much novelty}: A formative situation should be constructed so as to create a use context in which the participant can respond genuinely and appropriately to the prototype, moving past its status as a novel interaction or technology. The situation should encourage and allow for emergent \cite{Gaver2022EmergenceResearch} insights from participants about the intended context of use, that inform future design iteration and contribute to extensible design knowledge. 
    \item \textit{Shallow participant feedback}: Feedback elicited from participants in a formative situation should be rich, and speak to the situation itself, getting beyond surface, summative feedback such as "I liked the experience", that doesn't really go deeper to understand the ``WHY." As mentioned before, the designer can involve themself in the study (e.g. eliciting more detailed feedback, engaging in interviews) where appropriate in order to elicit more nuanced and rich feedback. 
\end{itemize}

\section{Reflections and Call to Action}
In this paper, we propose the use of formative situations as an additional type of RtD contribution, worthy of publication. We have outlined a proposed set of guidelines to help researchers report on such results.

Industry practitioners seek practical design knowledge and innovations from our research community. We believe they might find design insights about mid-fidelity prototypes gleaned from well-crafted formative situations more actionable than summative evaluations of non-commercially ready research prototypes. We as a RtD community may want to reassess the stakeholders for the artifacts we make. What is the purpose of the design knowledge we produce and why? Opening up a novel design space, or providing workable, practical design insights at the formative stages can be seen as a valuable contribution to the HCI community as such. Sometimes, it might make the most sense to just stop there. Other projects might warrant summative evaluations as well, but we argue that it is inappropriate to hold all research to this `gold standard' which may not even play to our community's strengths. After all, overly circumscribing design choices may obfuscate the underlying design insights that another researcher or industry practitioner could absorb and apply to a fully resolved research or commercial product. 

We hope this paper sparks a broader conversation about our community's bias toward summative evaluation. Rather than prematurely evaluating design artifacts before they are ready, we propose a shift in the community toward accepting formative evaluations of mid-fidelity prototypes as a standalone type of contribution. These contributions would not simply be considered `works-in-progress', but valid \textit{full} contributions to the field of HCI. Rather than racing to conduct summative evaluations of their prototypes to complete a publishable research cycle, designers could thoughtfully craft formative situations that offer useful knowledge to other researchers, as well to non-academic practitioners in industry and other venues who might use it in developing products. 
Research is all about providing the community with reusable, extensible knowledge. However, as it stands we all work within very dispersed research `islands', rarely exchanging ideas and taking up work that another research group or lab has made. The same research group that conceptualizes the work handles the prototype up until the resolution. We don't have to be and \textit{shouldn't} be islands---as this stymies innovation. We should be more freely and openly sharing and trading knowledge with other research groups at earlier stages, and more deliberately using that feedback to inform each other's work. Perhaps this type of knowledge contribution could help improve the translation and transition process between research and industry, by not over-constraining design knowledge into particular high fidelity prototypes and artificial situations crafted to support evaluative rigor. Overall, we assert that it is not always appropriate for every design and situation to conduct a summative evaluation. The list of criteria in this paper is by no means complete - we hope it becomes `community curated' in future, allowing others to give feedback and add to/adapt the criteria based on others' experiences.

We hope to see formative evaluations become accepted in the community as having rigor, and evaluated as such. In the future, we hope to both establish and evaluate a set of criteria for both researchers wishing to create formative situations of their own, and reviewers looking for guidance on how to assess this kind of work. We propose that summative user studies not be the only gold standard of HCI design research evaluations, but rather propose that we can evaluate design research artifacts at the appropriate design resolution, and still contribute valuable knowledge to the community. 

This paper gives researchers both the permission and structure within which to recognize and evaluate formative work. While the criteria are a proposed solution to the issues raised in the introduction, we would like to clarify that other sub-disciplines (and kinds of design work) within HCI might see different issues than us, and thus the criteria we propose may not fit their particular formative work. We believe that true empiricism is flexible \cite{Gaver2012WhatDesign}, and there exists a broad range of what can be considered rigorous and formative. We contribute this paper as a way to open up a conversation and orient the community to one of the many possible kinds of formative design knowledge that we've identified. We as an RtD community should be flexible and alert to the many different forms of empirical research that may exist, and recognize that future research papers might have other categorizations and definitions of very different rigorous, formative work that falls outside of what is defined within this paper. 
\section{Conclusion}
This paper presents formative situations as a method for generating design insights and knowledge at the mid-fidelity prototyping stage of the design process. Formative situations aim to elicit rich and subjective feedback from participants that has good ecological validity, and that can be taken up by others who intend to push forward in the related design space, whether in research or industry. We propose five criteria for the evaluation and assessment of formative situations. We propose that a formative situation should be well fitted to the RtD questions (C1), draw upon related best practices (C2), be of appropriate weight in proportion to the provisionality of the prototype (C3), allow designers an appropriate position within the process (C4), and allow participants an active role in the process (C5). In the future, we hope to show these criteria to various RtD experts in order to receive feedback regarding their use in practice, and unpack specific formative works at CHI that we see fitting these criteria. We offer this list as provisional and adaptable to community feedback. Ultimately, this paper aims to provide the community with a launching point for future discussion surrounding formative work in RtD.

\begin{acks}
A very deep thank you to both Elisa Mekler and Bill Gaver for their feedback on this work.
\end{acks}

\bibliographystyle{ACM-Reference-Format}
\bibliography{main}


\end{document}